\documentclass{iopart} 
\usepackage{iopams}              %  need boldface greek characters;  
\usepackage[dvips]{graphicx}     %  include EPS figures; 
\begin{document}
\jl{2} 
%   V    X    V    X    V    X    V    X    V    L    V    X    V    X    V    X
\title[Coherent and incoherent atomic scattering: Formalism and 
application]{Coherent and incoherent atomic scattering: Formalism and 
application to pionium interacting with matter}
\author[T A Heim, K Hencken, D Trautmann, and G Baur]% 
{T A Heim\dag, K Hencken\dag, D Trautmann\dag\ and G Baur\ddag}
\address{\dag\ Institut f\"ur Theoretische Physik, Universit\"at Basel,
4056 Basel, Switzerland}
\address{\ddag\ Institut f\"ur Kernphysik, Forschungszentrum J\"ulich,
52425 J\"ulich, Germany} 
\begin{abstract}
The experimental determination  of the lifetime of pionium provides a 
very important test on chiral perturbation theory. This quantity is
determined in the DIRAC experiment at CERN. In the analysis of this 
experiment, the breakup probabilities of pionium in matter are needed 
to high accuracy as a  theoretical input. We study in detail the 
influence of the target electrons.
They contribute through screening and incoherent effects. We use 
Dirac-Hartree-Fock-Slater 
wavefunctions in order to determine the corresponding form factors.
 We find that the inner-shell electrons contribute less than 
the weakly bound outer electrons. Furthermore, we establish a more
rigorous estimate for the magnitude of the contribution from the 
transverse current (magnetic terms thus far neglected in the 
calculations).
\end{abstract}
%\submitted 
%\maketitle
\section{Introduction}
In experiments such as atom--atom scattering, atom--electron
scattering,  nuclear scattering at high energy, 
or photo-production of $e^+e^-$ pairs on atoms
one faces the situation of complex systems
undergoing transitions between different internal states. For an 
atomic target, excitation affects the electrons individually. Thus
the ``form factor'' for the target-inelastic process takes the form 
of an \emph{incoherent} sum over all electrons, as opposed to the 
\emph{coherent} action of the electrons (and the nucleus) in the 
target-elastic case. From this observation it is immediately obvious 
that the target-inelastic cross section is proportional to $Z$, the 
number of electrons in the target atom, whereas the target-elastic 
process scales with $Z^2$. Since incoherent scattering off the excited 
target electrons increases the cross section from its value due 
to coherent scattering off the atom, the effect is sometimes referred 
to as ``anti-screening'' \cite{McGuire,Anholt}. At large momentum 
transfer the anti-screening correction can be accurately approximated 
by increasing the coherent scattering cross section by the factor 
$(1+1/Z)$, see e.g.~\cite{Wheeler,Sorensen,Voitkiv}.
We will demonstrate in this article, however, 
that this simple re-scaling argument is not sufficiently accurate 
in general. At the same time we will show how to obtain far superior 
results in a Dirac-Hartree-Fock-Slater approach with manageable
numerical effort.

As an application of the general formalism discussed in this paper, 
we put our focus on the particular situation pertaining to experiment 
DIRAC. This experiment, currently being performed at CERN 
\cite{nemenov}, aims at measuring the lifetime of pionium, i.e., a 
$\pi^+\pi^-$ pair forming a bound state as an exotic, hydrogen-like 
atom. While moving through matter, this system can breakup via 
electromagnetic interaction, 
or it can annihilate in the strong hadronic decay 
$\pi^+\pi^-\rightarrow\pi^0\pi^0$ if the pionium is in an $s$-state. 
The lifetime for this decay is related to the $\pi\pi$ scattering 
length which in turn has been calculated in the framework of chiral 
perturbation theory. Experiment DIRAC therefore provides a crucial 
test for a theoretical low-energy QCD prediction of pion--pion 
scattering. The pionium is formed from pions produced in the collision 
of a high energy proton beam on a (heavy element) target foil on 
condition that the two charged pions have small relative momentum. 
The pionium production rate can be inferred from the double inclusive
production cross section for $\pi^+\pi^-$-pairs without final state
Coulomb interaction. The kinematical conditions of the experiment 
(target thickness of 100 to 200~$\mu$m, compared to a predicted decay 
length in the lab of roughly 15 to 20~$\mu$m) imply annihilation of 
the pionium through strong interaction still within the target foil.  
However, the experiment does not measure the neutral pions. 
Instead, \emph{charged} pions are actually detected, and pions from 
electromagnetic breakup of ``atomic'' pairs are distinguished 
by small relative momentum and small angular separation from a large
background of accidental ``free'' pairs with arbitrary momentum and no
directional correlation. Comparing the observed ``ionized'' atomic 
pairs with the number of pionium atoms produced for a given target 
material and thickness, one can extract the lifetime for the hadronic 
decay. Thus very accurate cross sections for the electromagnetic 
interaction between pionium and the target material are needed as an 
input in the analysis of the experiment.

The paper is organized as follows: In \sref{formalism} we review the
well established formalism for one-photon exchange, applied e.g.\ in
electron--hadron scattering. In \sref{longwave} the cross sections 
coming from the transverse photons
are estimated with the help of the long-wavelength limit. 
The formulas for the evaluation of atomic form factors and scattering 
functions in the framework of Dirac-Hartree-Fock-Slater theory are 
derived in \sref{calcDHFS},
followed by an analysis of simple alternative models in \sref{models}.
Although the results presented in \sref{results} have been obtained in 
the context of experiment DIRAC, many of the conclusions drawn in 
\sref{conclusion} are also valid more generally in atomic scattering. 
\section{Formalism}
\label{formalism}
In \cite{HHHTV99} we applied the semiclassical approximation to calculate 
the coherent (target-elastic) cross section for pionium--atom scattering,
demonstrating that this part can indeed be determined with the desired
accuracy of 1\%. However, contributions not yet 
included in \cite{HHHTV99} need to be added to achieve an overall accuracy
at the 1\% level. Here we are interested in those processes, where the 
atom is excited together with the pionium. 
We will treat the problem within the PWBA. For our derivation we will
follow closely the standard formalism of electron scattering as
found, e.g., in \cite{deForestW66,HalzenM84,GreinerS95}.

\begin{figure}
\centering
\includegraphics[height=5cm]{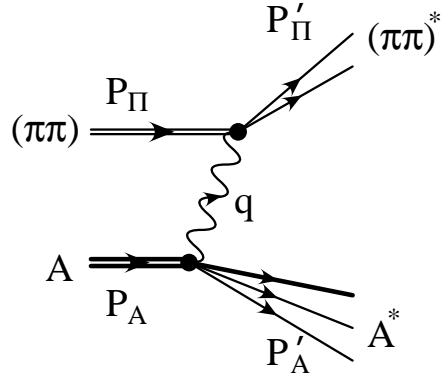}
\caption{The lowest order Feynman diagram for the simultaneous
excitation of projectile (pionium) and target (atom). The atomic 
momenta are $P_{\mathrm{A}}^{\vphantom{'}}$ and $P_{\mathrm{A}}'$ 
before and after the collision, those of the pionium 
$P_\Pi^{\vphantom{'}}$ and $P_\Pi'$. The momentum of the
exchanged photon is $q=P_\Pi'-P_\Pi^{\vphantom{'}}=
-(P_{\mathrm{A}}'-P_{\mathrm{A}}^{\vphantom{'}})$.}
\label{fig_feynman}
\end{figure}

We only calculate the lowest order result. The relevant Feynman
diagram is shown in \fref{fig_feynman}. The cross section for this
process is given by
\begin{equation}
\sigma = \frac{1}{4 I} \frac{1}{(2\pi)^2}
2 M_{\mathrm{A}} 2 M_\Pi \int\rmd^4q \frac{ (4\pi e^2)^2 
W^{\mu\nu}_{\mathrm{A}} W_{\mu\nu\Pi}}{(q^2)^2},
\end{equation}
where 
$I$ is the incoming flux, and 
$W^{\mu\nu}_{\mathrm{A}}$ and $W^{\mu\nu}_{\Pi}$ are the 
electromagnetic tensors describing the electromagnetic interaction of 
atom and pionium with the photon. As we are not interested in the 
specific final state of the atom, we can average over the possible
initial spin and sum over all final states and directions corresponding 
to a specific final state momentum $P'$. For the atom we have
\begin{equation}
\fl
W_{\mathrm{A}}^{\mu\nu} = \frac{1}{4\pi M_{\mathrm{A}}} 
\sum_X 
\left< 0, P_{\mathrm{A}} | J^{\mu\dag} | X, P_{\mathrm{A}}' \right>
\left< X, P_{\mathrm{A}}' | \vphantom{J^{\mu\dag}}
J^\nu | 0, P_{\mathrm{A}} \right>
 (2\pi)^4 \delta^4(P_{\mathrm{A}}-q-P_{\mathrm{A}}')
\end{equation}
and similarly for the pionium if its final state is not resolved either.

The electromagnetic tensor can only be a function of $P$ and
$P'$, or equivalently, of $P$ and $q$. Gauge invariance or
current conservation restricts the possible tensor structure of
$W^{\mu\nu}$ even more. It is a well known result that the
electromagnetic tensor in this case only depends on two scalar
functions $W_1$ and $W_2$ that are functions of $q^2$ and $Pq$
alone. The electromagnetic tensor is then given by
\begin{equation}
\fl
W^{\mu\nu} = \left(-g^{\mu\nu} + \frac{q^\mu q^\nu}{q^2}\right) 
W_1(q^2,Pq) + \left(P^\mu - \frac{Pq\; q^\mu}{q^2}\right)
\left(P^\nu - \frac{Pq\; q^\nu}{q^2}\right) \frac{W_2(q^2,Pq)}{M^2}. 
\label{eq_wmunu}
\end{equation}

Since the cross section depends on the product of the tensor for the atom
and the pionium, we calculate this product in terms of $W_1$ and $W_2$:
\begin{eqnarray}
\fl
W^{\mu\nu}_{\mathrm{A}} W_{\mu\nu\Pi} =
3  W_{1,\Pi} W_{1,\mathrm{A}}
+ \left( -1 + \frac{\Delta^2}{q^2}\right) W_{1,\Pi} W_{2,\mathrm{A}}
+ \left( -1 + \frac{\omega^2}{q^2}\right) W_{2,\Pi} W_{1,\mathrm{A}}
\nonumber\\ + \left( \gamma + 
\frac{\omega\Delta}{q^2}\right)^2 W_{2,\Pi} W_{2,\mathrm{A}},
\end{eqnarray}
where $\gamma$ is the (relative) Lorentz factor between atom and
pionium, $\Delta=-P_{\mathrm{A}} q /M_{\mathrm{A}}$, 
$\omega=P_\Pi q / M_\Pi$  are (minus) the energy
of the exchanged photon in the atom and pionium rest frame,
respectively. Following the argument of \cite{Afanasyev,HHHTV99} we 
expect the cross section to be dominated by the 
charge operator (termed ``scalar interaction'' in 
\cite{HHHTV99}). Thus we neglect at this point all terms containing a
$W_1$. We note also that $W_1$ vanishes in the elastic case for spin 0 
particles. However, the contribution to the cross 
section coming from $W_1$ will be discussed in 
\sref{longwave} below in the analysis of the transverse part of 
the current operator $\bi{j}$.
Keeping only the last term and assuming that the pre-factor is
dominated by $\gamma$ ($\gamma$ is in the range 15--20 in the DIRAC 
experiment) we get
\begin{equation}
W^{\mu\nu}_{\mathrm{A}} W_{\mu\nu\Pi} \approx \gamma^2 W_{2,\Pi} 
W_{2,\mathrm{A}}.
\end{equation}
Although this coincides with the naive estimate obtained by superficially
identifying the leading power of $\gamma$ (assumed to be dominant),
one should keep in mind that (as will be shown below) the range of 
$q^2$ starts at about $(\omega/\gamma)^2$ or $(\Delta/\gamma)^2$, 
where the last three pre-factors are of the same order. The magnitude of 
$\gamma$ should therefore not be mistaken as a justification
for neglecting the terms containing $W_1$. A critical assessment of 
the relative importance of $W_1$ as compared to $W_2$ is postponed 
until \sref{longwave}.

In our application to pionium--atom scattering
the masses $M_{\mathrm{A}}$ and $M_\Pi$ will be much larger 
than the momentum transfer $q$ of the photon. Therefore we will neglect
recoil effects on the atom and pionium. We can then identify $\Delta$
and $\omega$ as the excitation energy of the atom and pionium in
their respective rest frames.
Therefore $q_0$ and $q_z$ are fixed by the values of $\omega$ and
$\Delta$. In the following we will always denote the spatial part of the
photon momentum in the rest frame
of the atom by $\bi{k}$ and in the rest frame of the pionium
by $\bi{s}$. In the rest frame of the atom we then have
\begin{equation}
\eqalign{%
q_{0,\mathrm{A}} &= -\Delta\\
q_{z,\mathrm{A}} = k_z &= -\frac{\Delta}{\beta}-
\frac{\omega}{\gamma\beta},}
\end{equation}
and in the rest frame of the pionium
\begin{equation}
\eqalign{q_{0,\Pi} &= \omega\\ 
q_{z,\Pi} = s_z &= -\frac{\Delta}{\gamma\beta}-\frac{\omega}{\beta}.}
\end{equation}
Also $q^2$ is given by
\begin{equation}
q^2 = - \left( \frac{\Delta^2}{\beta^2\gamma^2} + 
\frac{\omega^2}{\beta^2\gamma^2} +
\frac{2\omega\Delta}{\beta^2\gamma} + q_\perp^2
\right) =: - \left( q_l^2 + q_\perp^2 \right).
\label{qsquare}
\end{equation}
Replacing the integration over $\rmd^4q=\frac{1}{\gamma\beta} \rmd\omega\
\rmd\Delta\ \rmd^2q_\perp$, we obtain 
\begin{equation}
\sigma = \int \rmd\omega \,\rmd\Delta\, \rmd^2q_\perp 
\frac{4\alpha^2}{\beta^2} \frac{W_{2,\Pi}(\omega,q^2) 
W_{2,\mathrm{A}}(\Delta,q^2)} {\left(q_l^2+q_\perp^2\right)^2}.
\label{sigmaW2W2}
\end{equation}
From \eref{qsquare} we note that $q_l^2$ may become negative if for example
the pionium is initially in an excited state and gets de-excited 
($\omega<0$) while the target atom gets excited ($\Delta>0$) from its 
ground state. For this process the cross
section \eref{sigmaW2W2} has a formal singularity, as pointed out recently
in \cite{VGG00}. Formally the divergence arises from integrating over the  
impact parameter all the way to infinity. Under the conditions of the 
experiment DIRAC, however, this formal divergence is negligible 
compared to the principal value of the cross section integral.

One of the advantages of this derivation is that the $W_2$ are scalar 
functions; we can therefore evaluate them in the respective rest frames,
even though the relative motion of atom and pionium is relativistic.
We now establish a relation between $W_2$ and the electromagnetic 
transition currents. As was already done in \cite{HHHTV99} we assume 
the charge operator 
to be the dominant contribution. In the atom's rest frame 
$W_2$ is related to the `00'-component of the tensor through 
\begin{equation}
W_{2,\mathrm{A}}=\frac{q^4}{k^4} W^{00}_{\mathrm{rf,A}},
\label{eq_w2w00}
\end{equation}
see, e.g., \cite{Walecka83,HenckenTB95}, with $k^2=\bi{k}^2=\Delta^2-q^2$.
Here we have added an index 
``rf'' as a reminder that the `00'-component in the rest
frame has to be taken. From the definition of $W^{\mu\nu}$ we get
\begin{equation}
\fl
W^{00}_{\mathrm{A}} = \frac{1}{4\pi M_{\mathrm{A}}} 
\sum_X 
\left< 0 | J^{0\dag}(q) | X, P_{\mathrm{A}}' \right>
\left< X, P_{\mathrm{A}}' | J^0(q) | 0 \right>
 (2\pi)^4 \delta^4(P_{\mathrm{A}}-q-P_{\mathrm{A}}').
\end{equation}
Rewriting this expression in terms of the (non-relativistic) density
operator and again neglecting recoil effects we find
\begin{eqnarray}
W^{00}_{\mathrm{A}} &= \sum_X \left< 0 | \rho(\bi{q}) | X, E_X \right>
\left< X, E_X | \rho(\bi{q}) | 0 \right> \delta(E_{0,\mathrm{A}}+
\Delta-E_X)\\ 
&=\sum_X \left|F_{X0,\mathrm{A}}(\bi{q})\right|^2 \delta(E_{0,\mathrm{A}}
+\Delta-E_X).
\end{eqnarray}
(Replacing the energy $E$ by the rest mass $M$ in these weakly bound
systems only introduces an error of the order $\alpha^2$). Finally we
obtain for the cross section:
\begin{eqnarray}
\sigma =& \int\rmd\omega\,\rmd\Delta\,\rmd^2q_\perp 
\frac{4\alpha^2}{\beta^2}\frac{q^4}{s^4 k^4} 
\left[ \sum_X \left|F_{X0,\mathrm{A}}(\bi{k})\right|^2
\delta(E_{0,\mathrm{A}}+\Delta-E_X)\right]\times\nonumber\\ &
\left[ \sum_{X'} \left|F_{X'0,\Pi}(\bi{s})\right|^2
\delta(E_{0,\Pi}+\omega-E_{X'})\right].
\label{eq_sigma}
\end{eqnarray}
As a verification of this formalism we reproduce the result of 
\cite{HHHTV99} for
target-elastic scattering. In this case $\Delta=0$ and
$k^2=q^2$, and the only possible final state is $X=0$ (assuming no
degeneracy of the ground state). We find 
\begin{equation}
\fl
\sigma = \int \rmd\omega \,\rmd^2q_\perp \frac{4\alpha^2}{\beta^2}
\frac{1}{s^4}
\left|F_{00,\mathrm{A}}(\bi{q})\right|^2
\left[ \sum_X \left|F_{X0,\Pi}(\bi{s})\right|^2
\delta(E_{0,\Pi}+\omega-E_X)\right],
\label{sigmael0}
\end{equation}
and $F_{00,\mathrm{A}}(q)$ is the elastic form factor of the atom.
This is identical to the equation derived in \cite{HHHTV99} in
the Coulomb gauge ($\bi{k}\cdot\bi{A} = 0$).

At this point we should add a few comments regarding the gauge invariance
and the factor $q^4/k^4$ (and likewise $q^4/s^4$) in \eref{eq_w2w00}:
In determining the general tensor structure of $W^{\mu\nu}$ in
\eref{eq_wmunu} the gauge invariance (or alternatively current 
conservation) was used. The magnitude of the component of $\bi{j}$ along 
the direction of $\bi{q}$ is then fixed by $\rho$. Only the transverse
parts of the current remain independent quantities. Our result therefore 
agrees with the one in Coulomb gauge, as in this gauge the component of 
$\bi{A}$ along the direction of $\bi{q}$ vanishes. An alternative 
approach might start directly from $W^{\mu\nu}$ without decomposition 
into $W_1$ and $W_2$. Assuming that in the rest frame of either the atom 
or the pionium $W^{00}$ dominates, one can approximate 
the product of the electromagnetic tensors by
\begin{equation}
W^{\mu\nu}_{\mathrm{A}} W_{\mu\nu\Pi} \approx \gamma^2 
 W_{\mathrm{rf},\mathrm{A}}^{00}W_{\mathrm{rf},\Pi}^{00}
\label{eq_w2w00b}
\end{equation}
where the factor $\gamma^2$ comes from the Lorentz transformation of one
tensor into the rest frame of the other. Now using \eref{eq_w2w00b} 
instead of \eref{eq_w2w00}, the cross section reads
\begin{eqnarray}
\sigma =& \int \rmd\omega \ \rmd\Delta\ \rmd^2q_\perp 
\frac{4\alpha^2}{\beta^2}\frac{1}{q^4} 
\left[ \sum_X \left|F_{X0,\mathrm{A}}(\bi{k})\right|^2
\delta(E_{0,\mathrm{A}}+\Delta-E_X)\right]\times\nonumber\\ &
\left[ \sum_{X'} \left|F_{X'0,\Pi}(\bi{s})\right|^2
\delta(E_{0,\Pi}+\omega-E_{X'})\right]. \label{sig_rf}
\end{eqnarray}
This differs from \eref{eq_sigma} by a factor $q^4/k^4\cdot q^4/s^4$.
In the elastic case \eref{sig_rf} corresponds to the result of 
\cite{HHHTV99} 
in the Lorentz gauge. However, the approximation \eref{eq_w2w00b} and
therefore also \eref{sig_rf} is not gauge 
invariant, whereas \eref{eq_sigma} is by
construction. We will therefore prefer in the following the form as
given in \eref{eq_sigma}. In \cite{HHHTV99} the difference between the 
two results was interpreted as an indicator  for the magnitude of the
``magnetic terms'', that is, the contribution proportional to $\bi{j}$.
In the next section we will estimate the contribution of
the transverse photons, making use of the long-wavelength limit. 
It remains to be seen, however, which one of the two possible schemes
(longitudinal/transverse decomposition versus scalar/magnetic terms)
will be better suited for explicit calculations.

In order to determine total cross sections we would like to simplify the
summation over all possible states. The expression \eref{eq_sigma}
depends on $\omega$ and $\Delta$ in two ways: Through the
energy-conserving delta functions and through the expression for $q^2$, 
where $q_l^2=\Delta^2/(\beta^2\gamma^2) +\omega^2/(\beta^2\gamma^2)
+(2\omega\Delta)/(\beta^2\gamma)$ depends on both $\omega$ and
$\Delta$. Replacing $\omega$ and $\Delta$ in $q_l$ by some average 
values $\omega_0$ and $\Delta_0$, we can perform the closure over all 
final states to get
\begin{equation}
\sigma = \int \rmd^2q_\perp \frac{4\alpha^2}{\beta^2}
\frac{q^4}{s^4 k^4} S_{\mathrm{inc,A}}(\bi{k}) S_{\mathrm{inc},\Pi}
(\bi{s})
\label{sigmasumrule}
\end{equation}
with $q^2, s^2$, and $k^2$ now the ones using $\omega_0$ and $\Delta_0$ 
and
\begin{eqnarray}
S_{\mathrm{inc,A}}(\bi{k})= \sum_X \left|F_{X0,\mathrm{A}}(\bi{k})
\right|^2 \\
S_{\mathrm{inc},\Pi}(\bi{s}) = \sum_{X'} \left|F_{X'0,\Pi}(\bi{s})
\right|^2.
\end{eqnarray}
In \sref{results} the dependence of the cross section on the 
choice of $\Delta_0$ and 
$\omega_0$ will be studied by varying both parameters 
over a reasonable range. 
For the atomic scattering function 
$S_{\mathrm{inc,A}}$, this study requires analyzing the contributions of
individual electron excitations shell by shell, since the binding energies
vary from some eV to several keV for the different shells.
\section{Contribution of transverse photons}
\label{longwave}
In \sref{formalism} we have only calculated the effect coming from the 
longitudinal photons (that is, coming from the charge operator).
Already in \cite{Afanasyev,HHHTV99} the effect of
the so-called ``magnetic interaction'', that is, the effect of the
current operator was estimated to be of the order of 1\%. As the
DIRAC experiment requires an accuracy of 1\%, these contributions need
to be considered more carefully. The part of the current operator 
$\bi{j}$ in the direction of $\bi{q}$ is already included in the above 
calculation. Therefore we need to study only the contribution coming 
from the transverse part of the current operator, that is, the effect 
coming from the transverse photons. Here we estimate them with the help 
of the long-wavelength approximation. Following \cite{Walecka83}, see 
also \cite{HenckenTB95}, $W_1$ and $W_2$ can be expressed in terms of 
the Coulomb and transverse electric (magnetic) matrix elements:
\begin{eqnarray}
W_1 = 2 \pi \left(|T^{\rme}|^2 + |T^{\mathrm{m}}|^2 \right)
\label{eq_w1temc} \\
W_2 = \frac{q^4}{s^4} 2 \pi \left( 2 |M^{\mathrm{C}}|^2 - \frac{s^2}{q^2} 
\left(|T^{\rme}|^2 + |T^{\mathrm{m}}|^2 \right) \right),
\label{eq_w2temc}
\end{eqnarray}
where we use the usual definitions for $M^{\mathrm{C}}$ and 
$T^{\rme,\mathrm{m}}$: $W^{00}=4\pi | M^{\mathrm{C}}|^2$ and
$W^{\lambda\lambda'}=\delta^{\lambda\lambda'} 2 \pi \left(
|T^{\rme}|^2 + |T^{\mathrm{m}}|^2 \right)$ with $\lambda,\lambda'$ 
denoting the two transverse 
directions. The $T^{\mathrm{m}}$ can be safely neglected in our
case. The long-wavelength limit relates $T^{\rme}$ to $M^{\mathrm{C}}$. 
In the multipole expansions 
\begin{equation}
T^{\rme}=\sum_{J\ge 1}\sum_M T^{\rme}_{JM},\qquad\qquad
M^{\mathrm{C}}=\sum_{J\ge 0}\sum_M M^{\mathrm{C}}_{JM}, 
\end{equation}
we have for 
a given multipole \cite{deForestW66,BlattW52}
\begin{equation}
T_{JM}^{\rme} \approx
\frac{\omega}{s} \left(\frac{J+1}{J}\right)^{1/2}
M_{JM}^{\mathrm{C}}, \qquad\mathrm{for\ } J\ge 1.
\label{TetoMC}
\end{equation}
In the case of the pionium only odd multipoles contribute and there is 
thus no $M_{00}^{\mathrm{C}}$ term. As higher multipoles are strongly 
suppressed we can safely set the factor $(J+1)/J$ to its maximum value at
2 (i.e.\ $J=1$) for the purpose of an analytical estimate for the upper 
limit of $|T^{\rme}|$. However, since our computer program 
already contains a multipole expansion, it is
of course straightforward to perform a more refined numerical calculation. 
For the atom, on the other hand, $M^{\mathrm{C}}_{00}\neq 0$;  
it is in fact the dominant contribution in the elastic process. 
Although generalizing the relation \eref{TetoMC} with the maximum factor 
$\sqrt{2}$ for all multipole orders still provides 
an upper limit for $T^{\rme}$, it may be a less useful over-estimate 
for the atom. 
The relation between $|M^{\mathrm{C}}|^2$ and $|T^{\rme}|^2$ is then
approximately
\begin{eqnarray}
|T^{\rme} (q)|^2 &\approx 2 \frac{\omega^2}{s^2} |M^{\mathrm{C}}(q)|^2 
\\ &\approx \frac{1}{2\pi} \frac{\omega^2}{s^2} W^{00}_{\mathrm{rf}}.
\label{TeW00}
\end{eqnarray}
One sees that the transverse photons contribute to both $W_1$ and $W_2$.
Denoting their contribution by $W_1^{\mathrm{T}}$ and 
$W_2^{\mathrm{T}}$, they can be expressed with the help of 
\eref{eq_w1temc}, \eref{eq_w2temc}, and \eref{TeW00} as
\begin{eqnarray}
W^{\mathrm{T}}_1 = \frac{\omega^2}{s^2} W^{00}_{\mathrm{rf}},\\
W^{\mathrm{T}}_2 = \frac{-q^2 \omega^2}{s^4} W^{00}_{\mathrm{rf}}.
\end{eqnarray}

First we look at the elastic case (for the atom). In this case we have
$\Delta=0$, $W_{1,\mathrm{A}}=0$ and 
$W^{00}_{\mathrm{rf,A}}=|F_{00,\mathrm{A}}(\bi{q})|^2$. The product
of the electromagnetic tensors is then
\begin{eqnarray}
W_{\mathrm{A}}^{\mu\nu} W_{\mu\nu\Pi} &= - W^{\mathrm{T}}_{1,\Pi} 
W_{2,\mathrm{A}} + \gamma^2 W^{\mathrm{T}}_{2,\Pi} W_{2,\mathrm{A}}\\
&= \frac{\omega^2}{s^2} \left( -1 + \gamma^2 \frac{-q^2}{s^2} \right) 
W^{00}_{\mathrm{rf},\Pi} \,|F_{00,\mathrm{A}}(\bi{q})|^2.
\end{eqnarray}
Again summing over all excited states of the pionium and neglecting the 
dependence on $\omega$ in $q^2$ and $s^2$ we get the total cross section
\begin{equation}
\sigma^{\mathrm{T}}_{\mathrm{el}} = \int \rmd^2q_\perp 
\frac{4\alpha^2}{\beta^2}\frac{\omega^2}{s^2} 
\left(-\frac{1}{\gamma^2 q^4} + \frac{1}{s^2 (-q^2)}\right)
S_{\mathrm{inc},\Pi}(\bi{s}) \,|F_{00,\mathrm{A}}(\bi{q})|^2.
\label{sigmaelT}
\end{equation}
We see that this cross section differs from \eref{sigmael0} by the
replacement
\begin{equation}
\frac{1}{s^4} \rightarrow 
\frac{\omega^2}{s^2}
\left(-\frac{1}{\gamma^2 q^4} + \frac{1}{s^2 (-q^2)}\right)
\end{equation}
The estimate for the reduction of these terms compared to the charge 
contribution can be seen easily in this equation, if one assumes that 
the dominant contributions come from the range of $q^2$ (and therefore 
also of $s^2$) of the order of $k_\Pi^2$, where $k_\Pi$ denotes the 
Bohr-momentum of the pionium, 
$k_\Pi=1/a_{\mathrm{Bohr},\Pi}\approx 136.566/a_{\mathrm{Bohr}}$. 
This momentum is of the order $\omega/\alpha$. Therefore the
factor $\omega^2/s^2$ would give a reduction of the order
$\alpha^2$, making this contribution completely negligible. 
However, a discussion of the relevant range of $q^2$ to be given
in \sref{results} will show that this estimate is too crude.

In the inelastic case (on the atom side) we can approximately
set $\Delta_0\approx 0$ again, as the atomic binding energy is small 
compared to the
other energies. Then the contribution from the transverse current of 
the atom will be suppressed even more than for the pionium, since the
ratio $\Delta/k$ is even smaller than $\omega/s$.
The only difference compared to \eref{sigmaelT} 
is then the replacement of $|F_{00,\mathrm{A}}(\bi{q})|^2$ by
$S_{\mathrm{inc,A}}(\bi{k})$. We get
\begin{equation}
\sigma^{\mathrm{T}}_{\mathrm{inel}} = 
\int \rmd^2q_\perp \frac{4\alpha^2}{\beta^2}
\frac{\omega^2}{s^2}
\left(-\frac{1}{\gamma^2 q^4} + \frac{1}{s^2 (-q^2)}\right)
S_{\mathrm{inc},\Pi}(\bi{s}) \,S_{\mathrm{inc,A}}(\bi{k}).
\label{sigmainelT}
\end{equation}
We will discuss the contribution to the cross section in \sref{results}.
\section{Dirac-Hartree-Fock-Slater model}
\label{calcDHFS}
We now turn to the question how to evaluate the form factors and 
scattering functions derived in the general formalism of the preceding 
sections. For our specific application of pionium scattering off atomic 
targets, we use the pionium form factors as described in \cite{HHHTV99}; 
here we shall only discuss the calculation of the \emph{atomic} form 
factors.

The purpose of this explicit calculation of atomic form factors and 
especially atomic scattering functions is twofold: Existing tables 
\cite{HubbellS} give no indication about the contributions from 
individual atomic shells, an information that is crucially needed to 
determine the appropriate average excitation energy in the closure 
approximation. Furthermore, since target atoms as heavy as Pt ($Z=78$) 
will be employed in the experiment DIRAC, the atomic 
structure is more appropriately treated with relativistic orbitals, 
at variance with the non-relativistic calculations in \cite{HubbellS}.
\subsection{Atomic ground state elastic form factors}
Within the framework of (Dirac-)Hartree-Fock-Slater theory, the atomic 
ground state wavefunction entering the expression for the form factor,
\begin{equation}
F_{00}(\bi{k}) = \langle\Psi_0 | \sum_{j=1}^Z \exp(\rmi\bi{k}\cdot
\bi{r}_j)| \Psi_0\rangle, 
\label{F00base}
\end{equation}
is given by a single Slater determinant constructed from products of 
independent particle orbitals,
\begin{equation}
\Psi_0 = \frac{1}{\sqrt{Z!}}\sum_{p}\,\mathrm{sign}(p)\,
\Phi_{p(1)}(\bi{r}_1)\cdots \Phi_{p(Z)}(\bi{r}_Z).
\end{equation}
Here $Z$ denotes the nuclear charge (and the number of electrons), while
$p$ denotes the permutations of orbital indices, and $\Phi_j$ signify
single particle orbitals.  Each of the $Z$ exponential terms $\exp(\rmi
\bi{k}\cdot\bi{r}_j)$ in \eref{F00base} 
acts as a one-particle operator. Orthogonality of 
the orbitals effectively cancels the summation over permutations
in the bra and ket vectors leading to
\begin{equation}
F_{00}(\bi{k})
= \sum_{j=1}^Z 
\langle\Phi_{j}(\bi{r}_j)|\exp(\rmi\bi{k}\cdot\bi{r}_j)|
\Phi_{j}(\bi{r}_j)\rangle .
\end{equation}
So far we have not considered any angular momentum coupling of the
independent-particle orbitals, that is, our $\Psi_0$ is determined by a
set of quantum numbers $(n_j,l_j,m_j),\,j=1\ldots Z$, without coupling to
a total $L$ (or $J$) and $M$. Furthermore we  
assumed the ground state wavefunction in the bra and ket
symbols to be identical. This need not actually be the case, as different
$z$-projections $M$ of the total angular momentum $J$ of the ground state
cannot be distinguished (without external fields). The angular momentum
coupled ground state wavefunction would then be obtained by summing and
averaging over $M$ and $M'$ in the bra and ket vectors, respectively.
Likewise on the level of independent particle labels $m_j$, the 
orthogonality argument applies strictly only to all orbitals except $j$, 
that is, we should distinguish between $m_j$ and $m_j'$ for the bra and 
for the ket vector, respectively. Since the one-particle operator does
not affect the spin part of the wavefunction, we should also insert an 
orthogonality factor for the spin orbitals in bra and ket, 
$\chi_j, \chi_j'$. Of course the principal and azimuthal 
quantum number $n_j$ and $l_j$ coincide
in the bra and in the ket symbol.

Expanding the exponential in spherical harmonics we find immediately 
\begin{eqnarray}
\fl
F_{00}(\bi{k}) = \sum_{j=1}^Z (2l_j+1)\delta_{\chi_j,\chi_j'}
\sum_{\lambda,\mu} \rmi^\lambda 
\sqrt{4\pi} Y_{\lambda,\mu}^*(\hat{k}) \sqrt{2\lambda+1}\left(
\begin{array}{ccc}l_j & l_j & \lambda \\ 0 & 0 & 0 \end{array}\right) 
\times \nonumber \\ (-1)^{m_j'}\left(\begin{array}{ccc}l_j & l_j & 
\lambda \\ m_j & -m_j' & \mu \end{array}\right)\,
\mathcal{R}_{jj}^{\lambda}(k),
\end{eqnarray}
with the radial form factor defined by
\begin{equation}
\mathcal{R}_{ij}^{\lambda}(k)=\int_0^\infty\rmd r\; r^2\,
R_{n_il_i}(r)\,j_\lambda(kr)\,R_{n_jl_j}(r),
\end{equation}
where the $R_{nl}(r)$ denote radial wavefunctions for the orbitals,
and $j_\lambda(kr)$ is a spherical Bessel function. 

In a next step averaging $|F_{00}(\bi{k})|^2$ over all directions 
$\hat{q}$ and using the orthogonality relation of the spherical 
harmonics yields
\begin{eqnarray}
\fl
|F_{00}(k)|^2 :=  
\frac{1}{4\pi}\int\rmd\hat{k}\,|F_{00}(\bi{k})|^2 \nonumber \\  \lo= 
\sum_{\lambda,\mu}(2\lambda+1)\left\{\sum_{j=1}^Z(-1)^{m_j'}(2l_j+1)
\delta_{\chi_j,\chi_j'}\left(
\begin{array}{ccc} l_j & l_j & \lambda \\ 0 & 0 & 0 \end{array}\right)
\right. \times\nonumber \\ \left.
\left(\begin{array}{ccc} l_j & l_j & \lambda \\ m_j & -m_j' & \mu 
\end{array}\right)\,\mathcal{R}_{jj}^{\lambda}(k)\right\}^2.
\label{F00DHFS}
\end{eqnarray}
Obviously all electrons contribute coherently to the form factor, as 
expected. Due to the first $3j$-symbol only even multipoles contribute 
to the sum.

In the $LS$-coupling scheme, the atomic ground state is characterized 
by a specific total $L$ and a (possibly averaged) total $M$. Instead of 
coupling the individual electrons' angular momenta to total $L$ and then 
averaging over $M$, we average directly over individual $m_j, m_j'$ for 
the orbitals occupied in accordance with the Pauli principle. This 
amounts to neglecting energy differences between fine structure levels. 
Hund's rules (see e.g.~\cite{Slater}) state that sub-shells 
are to be filled by adding as many electrons with different $m_j$ and 
the same spin projection as possible. The critical multiplicity is thus
that for a half filled sub-shell, $(2l+1)$. For half filled or completely 
filled sub-shells, Hund's rules imply that both $m_j$ and $m_j'$ run 
over all possible values from $-l_j$ to $+l_j$. Consequently the angular 
part for these spherical sub-shells reduces to selecting only monopole 
contributions. A completely filled sub-shell $(n_j,l_j)$ thus contributes 
\begin{displaymath}
2(2l_j+1)\mathcal{R}_{jj}^0(k)
\end{displaymath}
to the full form factor. (The leading factor 2 indicates 
the spin multiplicity.)

For sub-shells that are neither completely nor half filled the averaging
procedure yields different multiplicities depending on the multipole 
order. For the dominating monopole contribution, this factor is simply 
given by the occupation number of the sub-shell, whereas for higher 
multipoles this factor is proportional to the product of occupation 
number and the number of holes in a sub-shell with identical spin 
projections. For a given number of electrons in an open sub-shell we 
averaged the $m$-dependent part in \eref{F00DHFS} over all possible 
distributions of $m_j$ and $m_j'$ values.

Note that the coherent (elastic) form factor $F_{00}(k)$ derived above 
describes only the effect of scattering off the atomic electrons. The 
complete elastic form factor for the atom reads
\begin{equation}
F_{\mathrm{Atom}}(k)=Z-F_{00}(k),
\end{equation}
assuming a point-like nucleus.
\subsection{Atomic inelastic scattering functions}
Besides the elastic form factor $F_{00}(\bi{k})$ treated in the preceding 
section, we also need to consider the contributions due to excitations 
of the atomic electron cloud. Nuclear excitations will not be considered 
here because the much larger excitation energy required (typically on 
the order of MeV) exceeds the energy range relevant for our application 
to pionium--atom scattering. We demonstrated in \cite{HHHTV99} that 
deviations from a point-like nucleus are negligible for the 
electromagnetic processes considered here. Thus the nucleus' internal 
structure with its excited states is equally irrelevant as the 
experiment DIRAC cannot probe this structure. 
In analogy to the elastic form factor, a transition form factor is
written in the form 
\begin{equation}
F_{X0}(\bi{k}) = \langle\Psi_X | \sum_{j=1}^Z \exp(\rmi\bi{k}\cdot
\bi{r}_j) | \Psi_0\rangle,
\end{equation}
corresponding to scattering with excitation of the atomic electrons
from the ground state to some excited state $X$. A similar expression was
studied in \cite{BaurHT98} in the context of the equivalent photon 
approximation.
The total inelastic scattering function is defined as the
incoherent sum over all 
states $X$ other than the ground state
\begin{eqnarray}
S_{\mathrm{inc}}(\bi{k}) & =  \sum_{X\ne 0}|F_{X0}(\bi{k})|^2 \nonumber 
\\ & =  \sum_{\mathrm{all}\;X}|F_{X0}(\bi{k})|^2 - |F_{00}(\bi{k})|^2 
\nonumber \\
 & =  Z +\sum_{i=1}^Z\sum_{j\ne i}
\langle\Psi_0|\exp(\rmi\bi{k}\cdot[\bi{r}_j-
\bi{r}_i])|\Psi_0\rangle- |F_{00}(\bi{k})|^2 . \label{Sincbase}
\end{eqnarray}
Here we have added the ground state in order to exploit the completeness 
of the set of states $\{X\}$: Expanding the squared modulus of $F_{X0}$ 
introduced a second (primed) set of variables $\bi{r}_j'$ which has been 
removed again by virtue of the completeness of the set of states $\{X\}$. 
Furthermore we evaluated the sum over the diagonal terms $i=j$
separately, obtaining the term $Z$ (since $\exp(\rmi\bi{k}\cdot[\bi{r}_j-
\bi{r}_i])\equiv 1$ in this case).
Using the same Slater determinant wavefunctions for $\Psi_0$ as in the
preceding section, expanding the double sum corresponding to 
$|F_{00}(\bi{k})|^2$, and combining terms with the last sum, we find (in
terms of the single-electron orbitals)
\begin{equation}
S_{\mathrm{inc}}(\bi{k}) =
Z - \sum_{i=1}^Z\sum_{j=1}^Z |\langle\Phi_i|\exp(\rmi\bi{k}
\cdot\bi{r})|\Phi_j\rangle |^2.
\label{incohSum}
\end{equation}
The new terms required in the determination of 
$S_{\mathrm{inc}}$ are the matrix elements 
\begin{eqnarray}
\fl
\langle\Phi_i|\exp(\rmi\bi{k}\cdot\bi{r})|\Phi_j\rangle  = 
\delta_{\chi_i\chi_j}(-1)^{m_i}\sqrt{4\pi(2l_i+1)(2l_j+1)}
\sum_{\lambda,\mu}Y_{\lambda,\mu}^*(\hat{k}) 
\rmi^\lambda\sqrt{2\lambda+1}\times\nonumber\\ 
\left(\begin{array}{ccc}l_i & l_j & \lambda \\ 0 & 0 & 0
\end{array}\right)\left(\begin{array}{ccc}l_i & l_j & \lambda \\ 
-m_i & m_j & \mu\end{array}\right) \,\mathcal{R}_{ij}^{\lambda}(k),
\end{eqnarray}
some of which (namely, those with $i=j$) have already been used in 
calculating $F_{00}(\bi{k})$. As before, $\chi_i,\chi_j$ denote the spin 
projections. Averaging over the directions $\hat{k}$ we obtain immediately
\begin{eqnarray}
\fl
S_{\mathrm{inc}}(k) := \frac{1}{4\pi}\int\rmd\hat{k} \;
S_{\mathrm{inc}}(\bi{k}) \nonumber\\ \lo=  Z - \sum_{i=1}^Z\sum_{j=1}^Z
\delta_{\chi_i\chi_j}
(2l_i+1)(2l_j+1)\sum_\lambda(2\lambda+1)\left(
\begin{array}{ccc} l_i & l_j & \lambda \\ 0 & 0 & 0 \end{array}\right)^2
\times\nonumber \\ 
\left(\begin{array}{ccc} l_i & l_j & \lambda \\ m_i & -m_j & m_j-m_i 
\end{array}\right)^2
\left[\mathcal{R}_{ij}^{\lambda}(k)
\right]^2. \label{SincDHFS}
\end{eqnarray}
Considering again the \emph{non-averaged} incoherent scattering function 
in the special case of completely filled sub-shell $l_i, l_j$, the 
summations over $m_i$ and $m_j$ remove the angular dependence on 
$\hat{k}$. The filled sub-shells thus contribute 
\begin{displaymath}
2(2l_i+1)(2l_j+1)\sum_\lambda (2\lambda+1)\left(\begin{array}{ccc}
l_i & l_j & \lambda \\ 0 & 0 & 0\end{array}\right)^2 
\left[\mathcal{R}_{ij}^{\lambda}(k)\right]^2
\end{displaymath}
to the incoherent form factor, with a factor 2 for the spin
multiplicities in both sub-shells (rather than a factor 4, because
the cross terms between sub-shells with opposite spin projections
drop out due to the orthogonality of the spin orbitals).

Except for $Z=58$ (Ce, not of interest to us) all atoms with 
$Z\le 90$ have in their ground state only one sub-shell that is neither 
half nor completely filled \cite{PDB}. 
(In atoms with two open sub-shells, one of them has $l=0$.) 
Thus we need not consider the $m$-averaging
procedure for cases where $m_i$ and $m_j$ come from two different
partially filled sub-shells with $l_i>0$ and $l_j>0$. In the cases of our
interest averaging over $m_i$ and $m_j$ is then straightforward. 
The $m$-averaged contribution from two different sub-shells reads
\begin{displaymath}
\frac{1}{2}\nu_i\nu_j\sum_\lambda (2\lambda+1)
\left(\begin{array}{ccc}l_i & l_j &
\lambda \\ 0 & 0 & 0\end{array}\right)^2 
\left[\mathcal{R}_{ij}^{\lambda}(k)\right]^2
\end{displaymath}
where $\nu_i$ and $\nu_j$ denote the occupation numbers of the two 
sub-shells and the factor $1/2$ stems from the orthogonality of spin 
orbitals. If $i$ and $j$ both refer to the same sub-shell the 
multiplicity factor is slightly more complicated, depending on whether 
the sub-shell is less than, or more than, half filled. For this limited 
number of cases, we again determined the $m$-dependent part by suitably 
averaging over all possible distributions of $m$ values in a partially 
filled sub-shell.
\section{Some other models}
\label{models}
For comparison we briefly discuss in this section some simple 
alternative models for evaluating the coherent and incoherent 
form factors, $F_{00}(k)$ and $S_{\mathrm{inc}}(k)$, respectively, 
for application in complex atomic scattering. Specifically, we
will use analytical screening models \cite{Salvat,Moliere} to derive the
elastic form factors. 
The inelastic scattering functions can then be obtained
from the elastic form factors either in the no-correlation limit, or 
from an argument due to Heisenberg and based on the Thomas-Fermi model.

The simplest possible model to describe the effect of incoherent 
scattering off the atom's electrons would merely divide the cross 
section for coherent scattering (scaling with $Z^2$)  by $Z$, on the 
grounds that everything remains the same except that each electron 
contributes incoherently to the cross section (complete 
``anti-screening''). We found that this approach underestimates the 
incoherent scattering cross section by as much as 50\%. For typical 
targets like Ti or Ni this implies an error of roughly 2\% in the 
target-inclusive cross section, clearly beyond the required limit 
of 1\%.
\subsection{Elastic form factors}
In order to simplify the atomic structure calculation, one might use the
Thomas-Fermi model to replace the density $\rho(\bi{r})$ occurring in
\begin{equation}
F_{00}(\bi{k}) = \int\rmd^3 r\,\rho(\bi{r})\exp(\rmi\bi{k}\cdot\bi{r}).
\end{equation}
Expressing the 
potential due to the charge distribution of the electrons in the form
\begin{equation}
V(r) = -\frac{Z}{r}\chi(r),
\end{equation}
the corresponding charge distribution is given by the second derivative of
the screening function:
\begin{equation}
\rho(r)=\frac{Z}{4\pi r}\chi''(r).
\end{equation}
Here and in the following the prime denotes differentiation with respect 
to $r$. For a spherical charge distribution the coherent form factor 
reduces to the monopole term. 
Using Moli\`ere's \cite{Moliere} parameterization for $\chi$,
\begin{eqnarray}
\chi(r) =  \sum_{i=1}^3 B_i\exp(-\beta_i r/b); \\
  B_1=0.1;\; B_2=0.55;\; B_3=0.35; \\
  \beta_1=6.0;\;\beta_2=1.2;\;\beta_3=0.3;
\end{eqnarray}
with $b=a_{\mathrm{Bohr}}(9\pi^2/128)^{1/3}\,Z^{-1/3}$, the coherent form
factor reads 
\begin{equation}
F_{00}(k) = Z\sum_{i=1}^3 B_i\frac{1}{1+(b\;k/\beta_i)^2}.
\label{Fmol}
\end{equation}
for the electronic part, and again $F_{\mathrm{Atom}}(k) = Z-F_{00}(k)$.
The same analytical form with different parameters $B_i$ and $\beta_i$ 
determined by fitting expectation values of powers of $r$ to exact 
Dirac-Hartree-Fock-Slater results
is used in \cite{Salvat} where the fitting parameters for all $Z\le 92$
may be found as well.
\subsection{Inelastic form factors: No-correlation limit}
Inserting the elastic form factor of the previous subsection into
\eref{Sincbase} we are left with the evaluation of
\begin{equation}
\fl
\sum_{i=1}^Z
\sum_{j\ne i}\langle \Psi_0|\exp(\rmi\bi{k}\cdot[\bi{r}_j-\bi{r}_i]|
\Psi_0\rangle = Z(Z-1)\int\rmd^3 r\,\rmd^3 r'\,N_2(\bi{r},\bi{r}')
\exp(\rmi\bi{k}\cdot[\bi{r}-\bi{r}'])
\end{equation}
where we have already integrated over all variables not pertaining to
the orbitals $i$ and $j$. Here the function $N_2(\bi{r},\bi{r}')$ 
describes the probability of finding any two of the
properly anti-symmetrized electrons at positions 
$\bi{r}$ and $\bi{r}'$. The \emph{no-correlation limit}
now replaces this two-particle density by the product of
the single-particle probabilities $\rho(\bi{r})/Z$ and $\rho(\bi{r}')/Z$. 
The integral in the last expression then reduces 
to $|F_{00}(\bi{k})/Z|^2$, i.e., the square of the
elastic form factor normalized per electron. As there are $Z(Z-1)$ such
terms in the double sum over $i,j$, we finally find
\begin{eqnarray}
S_{\mathrm{inc}}(\bi{k})=
 Z- |F_{00}(\bi{k})|^2/Z.
\label{nocorr}
\end{eqnarray}
Using the elastic form factor of the previous subsection, this result 
provides a simple expression for $S_{\mathrm{inc}}$. However, we have 
made here the crucial assumption that there is no correlation between
the electrons in different orbitals. In a single-particle picture this
amounts to assuming that all single-particle states are available for all
electrons simultaneously. 
Pauli blocking, i.e., the fact that 
due to the Pauli exclusion principle some states $X$ cannot be excited 
for a given electron because they are occupied by other electrons, 
is then disregarded completely. Since in this limit the (incoherent)
summation over $X$ also includes
the Pauli blocked states, the expression \eref{nocorr} clearly 
overestimates  the correct scattering function. 
\subsection{Thomas-Fermi model for incoherent scattering}
Following Heisenberg \cite{Heis} we can find a simple expression for the
\emph{incoherent} atomic form factor, going beyond the no-correlation 
limit but remaining in the spirit of the Thomas-Fermi model (see also 
\cite{Tsai}). Expanding the squared modulus in the incoherent sum 
\eref{incohSum} we write
\begin{equation}
S_{\mathrm{inc}}(\bi{k})=Z - \int\rmd^3 r\,\rmd^3 r'\,\exp(
\rmi\bi{k}\cdot(\bi{r}-\bi{r}'))|\sum_{j=1}^Z\Phi_j^*(\bi{r})
\Phi_j(\bi{r}')|^2. \label{Heis12}
\end{equation}
In the Thomas-Fermi model, the density is related to the volume of a 
sphere in momentum space, 
\begin{equation}
\sum_{j=1}^Z |\Phi_j(\bi{r})|^2=\frac{2}{(2\pi)^3}\int_{\kappa\le 
k_{\mathrm{F}}(\bi{r})}\llap{$\rmd^3 \kappa$}
= \frac{1}{3\pi^2} [2|V(r)|]^{3/2}.
\end{equation}
Heisenberg generalized this expression to obtain the two-particle density 
as an integral in momentum space (with $\bi{r}_1=(\bi{r}+\bi{r}')/2$)
\begin{equation}
\sum_{j=1}^Z \Phi_j^*(\bi{r})\Phi_j(\bi{r}')=\frac{2}{(2\pi)^3}
\int_{\kappa\le k_{\mathrm{F}}(\bi{r}_1)}\llap{$\rmd^3 \kappa$}
\,\exp(\rmi\bkappa\cdot(\bi{r}-\bi{r}'))
\end{equation}
which reduces to the Thomas-Fermi expression for $\bi{r}=\bi{r}'$. 
\eref{Heis12} now contains additional six-fold integration over $\rmd^3
\kappa$ and $\rmd^3 \kappa'$. Performing these integrations as well as 
the integration over $(\bi{r}-\bi{r}')$, we find (following
Heisenberg's argument about the common volume of two intersecting spheres
in momentum space, and paying attention to the spin multiplicity)
\begin{equation}
\fl
S_{\mathrm{inc}}(\bi{k})  =  
Z - \frac{4}{3\pi}\int_0^{r_0}\rmd r_1\, r_1^2
(\sqrt{2 V(r_1)}-k/2)^2(\sqrt{2 V(r_1)}+k/4),
\end{equation}
where the integration over $\bi{r}_1$ is restricted
to the region of coordinate space with $k_{\mathrm{F}}(\bi{r}_1)\ge k/2$ 
since otherwise the two spheres in momentum space do not overlap.  

Using $V(r)=(Z/r)\chi(r)$  the incoherent form factor turns into
\begin{equation}
\fl
S_{\mathrm{inc}}(k) =  
Z 
-\frac{4}{3\pi}\int_0^{r_0}\rmd r\, \sqrt{r} [2Z\chi(r)]^{3/2}
+\frac{2}{\pi}Z\,k\int_0^{r_0}\rmd r\;r\chi(r)
- \frac{1}{36\pi}(k\, r_0)^3 
\label{int32}
\end{equation}
where the upper limit of integration $r_0$ must satisfy
\begin{equation}
\frac{Z}{r_0}\chi(r_0) = \frac{1}{8}k^2.
\end{equation}

Within the frame work of the Thomas-Fermi model, the screening function
satisfies the differential equation
\begin{equation}
\frac{Z}{r}\frac{\rmd^2}{\rmd r^2}\chi(r) = 
\frac{4}{3\pi}[ (2Z/r)\chi(r)]^{3/2}; \qquad \chi(0)=1,
\end{equation}
thus enabling us to replace the first integral in \eref{int32}. The 
second derivative of $\chi$ can then be removed by integrating by parts, 
yielding
\begin{equation}
\fl
S_{\mathrm{inc}}(k) = 
- Zr_0\chi'(r_0) 
 +\frac{2}{\pi}Z\,k\int_0^{r_0}\rmd r\;r\chi(r)
+\frac{1}{8}r_0k^2 - \frac{1}{36\pi}(k\, r_0)^3.
\label{intTF}
\end{equation}
From \eref{int32} and \eref{intTF} we note that this simple model does
not reproduce the correct limit as $k\rightarrow 0$: In this limit 
$r_0\rightarrow\infty$, and the integral in \eref{intTF} assumes a 
constant non-zero value. Thus $S_{\mathrm{inc}}$ grows \emph{linearly} 
with $k$. By contrast the expression in the no-correlation limit 
(containing $F_{00}(k)$) grows with $k^2$, as does the
Hartree-Fock-Slater result derived in the preceding section (since the
term linear in $k$ drops out of the expansion in \eref{Sincbase} due to
the symmetry under interchange $i\leftrightarrow j$).
\section{Numerical method and results}
\label{results}
In our application to scattering of pionium on atomic targets, we use
hydrogenic wavefunctions for the pionium system $\pi^+\pi^-$. The 
corresponding form factors $F_{00,\Pi}$ and $S_{\mathrm{inc,}\Pi}$ 
are evaluated analytically as described in \cite{HHHTV99}. 

\Sref{calcDHFS} provides expressions for the evaluation of coherent
atomic form factors and incoherent scattering functions in the framework
of the Dirac-Hartree-Fock-Slater formalism. In our calculations we start
from a simple analytical charge distribution for the electrons as given
e.g.~in \cite{Salvat}  or similarly in \cite{Moliere}. 
Taking the electronic structure of the elements from \cite{PDB}
 we solve either the Schr\"odinger or the Dirac radial equation for each 
occupied orbital, treating exchange effects by Latter's approximation 
(see e.g.~\cite{Salvat}). The resulting charge density is
then used to obtain improved radial orbitals, iterating the process until
self-consistency is reached. Even for heavy elements with electrons in
some twenty different orbitals and requiring several ten iterations, this 
calculation is readily performed with the help of the program package 
RADIAL \cite{radial}. 

Using these orbitals we evaluate the radial integrals
in \eref{F00DHFS} and \eref{SincDHFS} on a reasonably dense mesh of $k$ 
values with the help of an integration routine developed for integrals 
containing spherical Bessel functions and powers \cite{BINS}. To this end, 
the numerical solutions for the orbitals obtained on a grid of $r$ values
are replaced by piecewise splines. The angular parts for the partially
filled sub-shells are determined by averaging over all distributions of
magnetic quantum numbers $m$ in accordance with the Pauli principle, 
as described in \sref{calcDHFS}.

\begin{figure}
\centering\includegraphics[width=10cm]{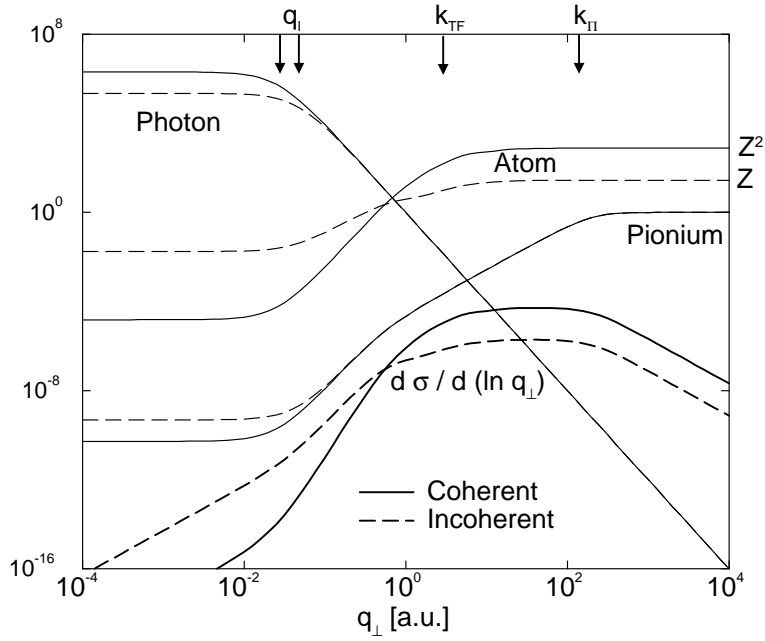}
\caption{Various contributions to the integrand for the cross section vs.
$q_\perp$, on a log-log scale. To compensate for the logarithmic 
$q_\perp$-axis, the integrand is represented as 
$\rmd\sigma/\rmd\ln q_\perp$. 
The individual curves are labelled in the figure.
The cross sections correspond to target-elastic 
(solid lines) and target-inelastic (dashed lines) pionium scattering off 
Ni ($Z=28$) at a projectile energy
$E=5$~GeV, summed over all final states of the pionium. 
The arrows at the upper edge indicate the relevant momentum scales.
See text for details.}
\label{fig_contrib}
\end{figure}

Let us first investigate the range of $q_\perp$ relevant for the cross 
section as given by \eref{sigmaW2W2} or \eref{sigmasumrule}.
\Fref{fig_contrib} demonstrates the interplay of different momentum 
scales associated with the photon, with the atom, and with the pionium, 
respectively.
The figure shows the integrand from \eref{sigmaW2W2} in singly 
differential form: 
\begin{equation}
\frac{\rmd\sigma}{\rmd q_\perp} = 
2\pi q_\perp\frac{4\alpha^2}{\beta^2}\,
(\mathrm{Photon})\times (\mathrm{Atom})\times (\mathrm{Pionium}),
\end{equation}
together with its decomposition into photon, atom
and pionium parts (cf.~\eref{qsquare}):
\begin{eqnarray}
\fl
(\mathrm{Photon})=\frac{1}{q^4}=\frac{1}{(q_l^2+q_\perp^2)^2}, \\
\fl
(\mathrm{Atom})=
W_{2,\mathrm{A}}=\cases{|F_{00,\mathrm{A}}(q)|^2 & for coherent
scattering, \\ (q^4/k^4)S_{\mathrm{inc,A}}(k) & for incoherent 
scattering,} \\
\fl
(\mathrm{Pionium})=
W_{2,\Pi}= (q^4/s^4)S_{\mathrm{inc},\Pi}(s).
\end{eqnarray}
The solid lines refer to the total cross section (i.e.,
summed over all pionium final states) for target-elastic 
scattering off Ni ($Z=28$) for pionium in its ground state. 
The projectile energy is 5~GeV. The dashed lines correspond to the same
setting for the target-inelastic process. Here the integrations 
over $\omega$ and $\Delta$ in \eref{sigmaW2W2} have been
replaced by setting an average excitation energy for the pionium at 
$\omega_0=1.858$~keV (ground state binding energy), and for the atom we 
set $\Delta_0=0$ (target-elastic) and $\Delta_0=$100~eV 
(target-inelastic), respectively.

The arrows indicate the relevant momentum scales: $q_l$ for the photon, 
$k_{\mathrm{TF}}=Z^{1/3}/a_{\mathrm{Bohr}}$ for the atom, 
and $k_{\Pi}$ for the pionium.
The arrow on the left under the label ``$q_l$'' corresponds to 
target-elastic scattering, while the one on the right corresponds
to incoherent
scattering (with non-zero $\Delta_0$ and thus with a larger $q_l$). 
For $q_\perp\ll q_l$, the photon momentum is essentially given by the 
constant $q_l$. When $q_\perp\gg q_l$, on the other
hand, this part displays a $1/q_\perp^4$ behavior.
The atomic part shows an increase between $q_l$
and several inverse Bohr radii (indicated by $k_{\mathrm{TF}}$).
As expected, $|F_{00,\mathrm{A}}(k)|^2$ grows roughly with $q_\perp^4$,
whereas $S_{\mathrm{inc,A}}(k)$ grows only with $q_\perp^2$. At $q_\perp
\approx 5k_{\mathrm{TF}}$ the atomic part reaches its asymptotic value
($Z^2$ or $Z$, respectively). In this regime the pionium part only just
starts to contribute appreciably. It grows quadratically with $q_\perp$
to saturate at a few multiples of the pionium scale $k_\Pi$.
The product of the three factors clearly demonstrates that the
main contributions to the cross sections come from the range of $q_\perp$
between $k_{\mathrm{TF}}$ and $k_\Pi$. 

In \fref{fig_contrib} we set $\omega_0=E_{\mathrm{bind},\Pi}$ for the
pionium, as well as a non-zero value for $\Delta_0$ in the case of 
incoherent scattering. The specific choice for these average excitation 
energies is guided by the following observations. For coherent 
(target-elastic) scattering $\Delta_0\equiv 0$, and any ambiguity in 
calculating total cross sections from \eref{sigmasumrule} is limited to 
the choice of an appropriate $\omega_0$. From a comparison 
\cite{HHHTV99} of total cross sections for coherent scattering in the 
closure approximation with the ``exact'' total cross sections obtained
by accumulating partial cross sections for bound-bound and bound-free
pionium transitions (summation/integration over all final states), we 
note that bound-bound transitions (excitation and de-excitation) account 
for the major part of the total cross section. Typically, breakup 
(ionization) accounts for some 30\%--40\% of the total cross section in 
the ground state, decreasing roughly by a factor $n^2$ for pionium in 
the initial state $(n,l)$. Furthermore, most of the breakup cross 
section from a given initial state comes from the range of continuum 
energies from 0 to about $E_{\mathrm{bind},\Pi}$ above the continuum 
threshold. Therefore the average energy difference between initial and 
all final states---weighted by their contribution to the total cross 
section---is of the order of the binding energy $E_{\mathrm{bind},\Pi}$. 

\begin{figure}
\centering\includegraphics[height=5cm]{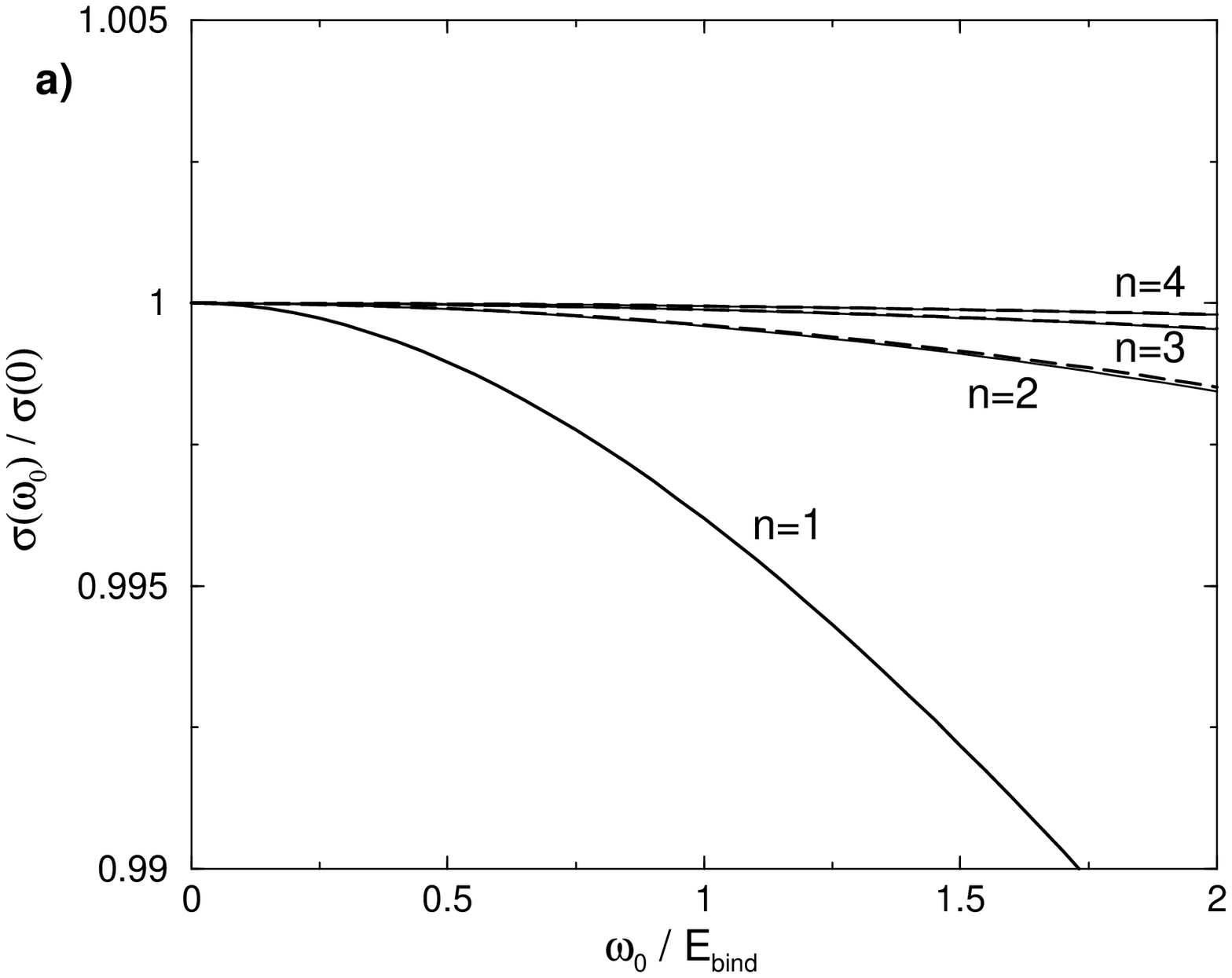}%
\qquad\includegraphics[height=5cm]{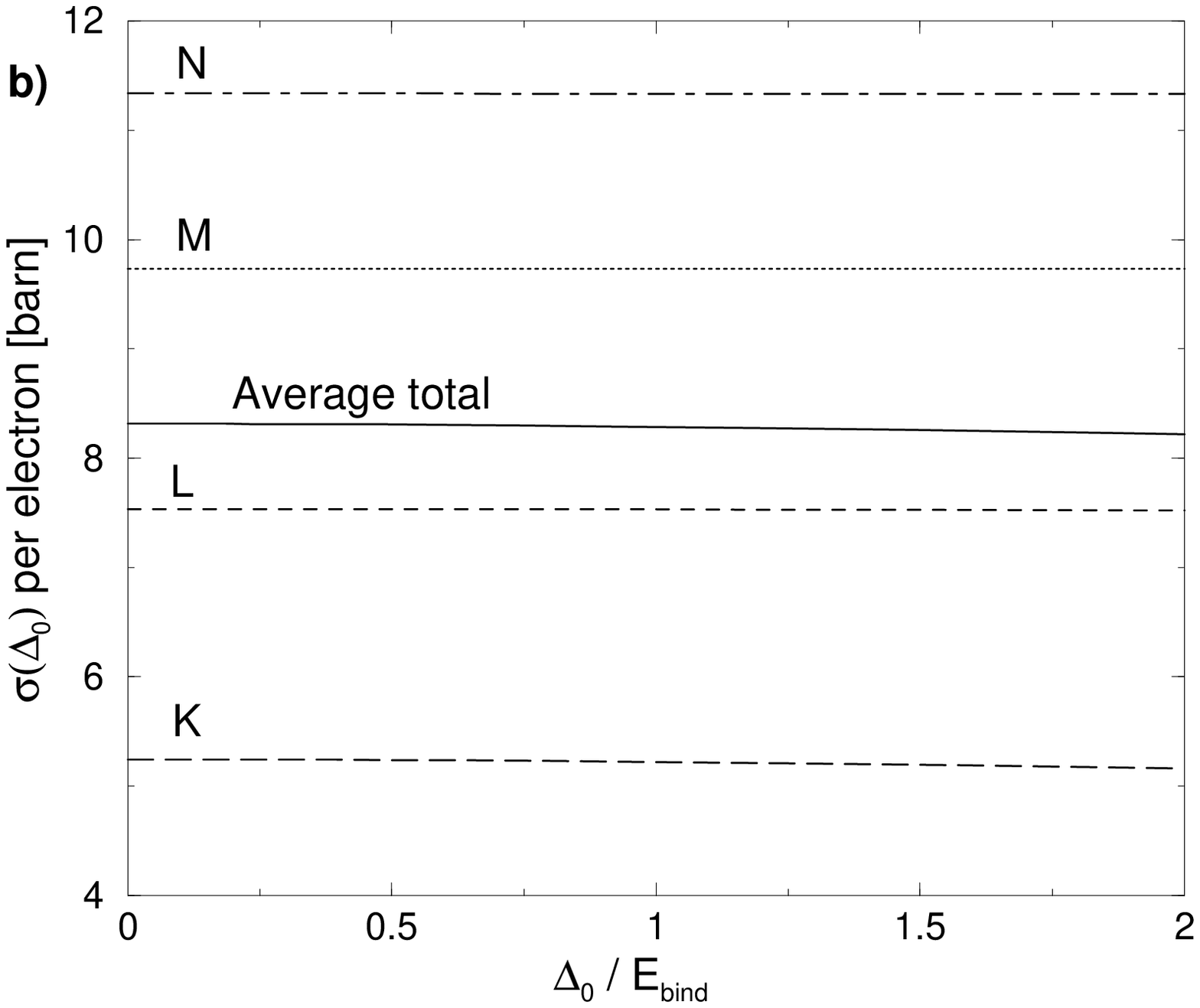}
\caption{\textbf{a)} Closure cross sections (normalized to $\omega_0=0$) 
versus $\omega_0$ (in units of $E_{\mathrm{bind},\Pi}$) for 
\emph{coherent} scattering of pionium in various initial states as 
indicated in the figure. Target material is Ni, projectile energy is 
5~GeV. Solid lines: initial $s$ states; dashed lines: initial $p$ 
states of the pionium. \textbf{b)} Contributions of individual target 
electrons to the closure cross section for \emph{incoherent} scattering 
of pionium in its ground state, plotted versus average atomic excitation 
energy $\Delta_0$ (normalized to the average binding energy of the 
respective atomic shells, ranging from some 10~eV for the $N$-shell to 
$\sim5$~keV for the $K$-electrons). Also shown is the average 
contribution of all shells (total incoherent cross section divided by $Z$).
Target material is Ti, projectile energy is 5~GeV.}
\label{sigvse}
\end{figure}
\Fref{sigvse}a) shows the total cross section for elastic scattering
in the closure approximation as a function of $\omega_0$ (in units of 
$E_{\mathrm{bind},\Pi}$). The total cross sections have been normalized 
to their
values at $\omega_0=0$. As can be seen from the figure, only the ground
state total cross section varies appreciably over a reasonable range of
$\omega_0$ values. We also find \cite{HHHTV99}
that the closure cross sections at 
$\omega_0=E_{\mathrm{bind},\Pi}$ coincide with the converged accumulated
partial cross sections. We therefore set $\omega_0=E_{\mathrm{bind},\Pi}$
in the following. 

In frame b) of the same figure we show the dependence on
the average atomic excitation energy $\Delta_0$ for 
the target-inelastic scattering process. 
Here the pionium energy has been fixed at $\omega_0\approx 1.858$~keV.
Since we calculate $S_{\mathrm{inc}}(k)$ using individual orbitals, we 
can easily determine the contributions resolved with respect to atomic
shells, or even per individual electron as in \fref{sigvse}b). 
The tabulated values for the  atomic incoherent scattering functions
\cite{HubbellS} cannot serve to study the closure approximation's 
sensitivity to the atomic excitation energy. When varying $\Delta_0$
from 0 to several keV, it is important to distinguish whether this 
excitation energy is transferred to an electron in the $K$ shell or to
one of the outer electrons. In the latter case, the atomic electron
is excited into a high energy continuum state. This process contributes
very little to the incoherent cross section which---like the pionium's
cross section---is dominated by excitation rather than by ionization. 
The average excitation energy $\Delta_0$ for each curve in \fref{sigvse}b)
has been normalized to the binding energy of the individual 
shells (averaged over sub-shells). 
We note that the individual
contributions are roughly proportional to the principal quantum number of 
each electron. Combined with the fact that the outer shells typically 
accommodate many more electrons than the inner 
shells, we find that the target-inelastic (incoherent)
scattering process is clearly dominated by the loosely bound outer
electrons. At the same time \fref{sigvse}b) also demonstrates that the 
incoherent scattering cross section is almost independent of $\Delta_0$.
Thus we may safely set $\Delta_0=0$ in our calculation.

\begin{figure}
\centering\includegraphics[width=8cm]{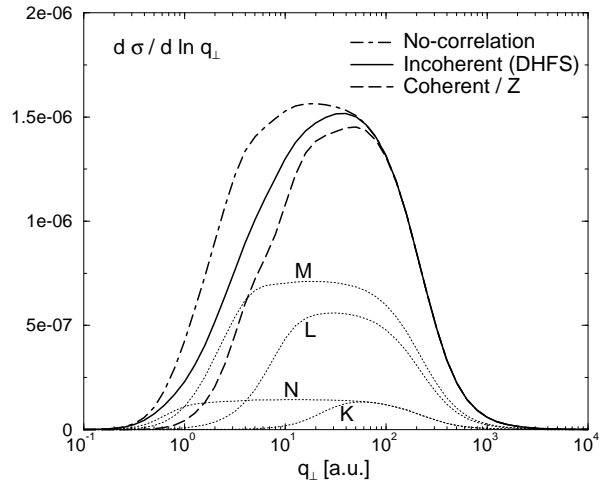}
\caption{Integrand for incoherent (target-inelastic) cross section of
ground state pionium scattering off Ti at energy 5~GeV, versus $q_\perp$
(on a log-scale). 
The areas under the curves yield the respective cross sections. 
Calculations with $\omega_0=1.858$~keV and 
$\Delta_0=0$. The contributions from the individual shells (dotted lines)
are labelled in the figure.}
\label{intbyshell}
\end{figure}
As a verification and further illustration of these findings, 
\fref{intbyshell} displays $\rmd\sigma_{\mathrm{inc}}/\rmd\ln q_\perp$
for ground state pionium scattering incoherently off Ti ($Z=22$) at 5~GeV
projectile energy. The solid line corresponds to the integrand for 
incoherent scattering \eref{sigmasumrule} calculated using 
\eref{SincDHFS}. 
In these calculations we set $\omega_0=1.858$~keV, the binding energy of
the pionium, and $\Delta_0=0$. As can be seen from the dashed 
line in the figure, the simplest approximation consisting 
of scaling the target-elastic (coherent) cross section by $1/Z$ clearly 
underestimates the correct result. At the same time the dash-dotted 
(chain) curve shows that the 
approximation using the coherent form factor \eref{F00DHFS} in the 
no-correlation limit obviously overestimates the correct result 
by an even  larger amount.  In more detail we find that the ratio 
$(\sigma_{\mathrm{coh}}/Z):\sigma_{\mathrm{inc}}$ for pionium initially 
in the ground state amounts to 0.49 (!) for a Li target, 0.62 (Al), 
0.66 (Ti), 0.73 (Ni), 0.72 (Mo), and 0.74 (Pt). For pionium initially 
in $2s$ or $3s$ states, these ratios range from 0.65 (Li) to 0.80 (Pt) 
for these targets relevant in the experiment. For the lighter targets, 
these huge uncertainties of the incoherent part result in considerable 
errors in the inclusive cross sections: The ratio 
\begin{displaymath}
\frac{\sigma_{\mathrm{coh}}(1+1/Z)}{\sigma_{\mathrm{coh}}+
\sigma_{\mathrm{inc}}} 
\end{displaymath}
amounts to 0.795 (Li), 0.958 (Al), 0.978 (Ti), 0.987 (Ni), 0.991 (Mo), 
and 0.996 (Pt). Thus only for the heavy targets the required accuracy of 
1\% can be attained with such a crude approximation for the incoherent 
part.

Further in \fref{intbyshell}, the dotted lines show the contributions 
to the incoherent
scattering cross section resolved according to atomic shells. A direct
comparison of the areas enclosed between the individual lines and the
abscissa demonstrates that the ten $M$-electrons dominate the cross 
sections, followed by the eight $L$-electrons. Also, the two 
$N$-electrons contribute considerably more strongly than the 
$K$-electrons, whose influence is limited to rather large $q_\perp$ as 
expected.

\begin{figure}
\centering\includegraphics[width=8cm]{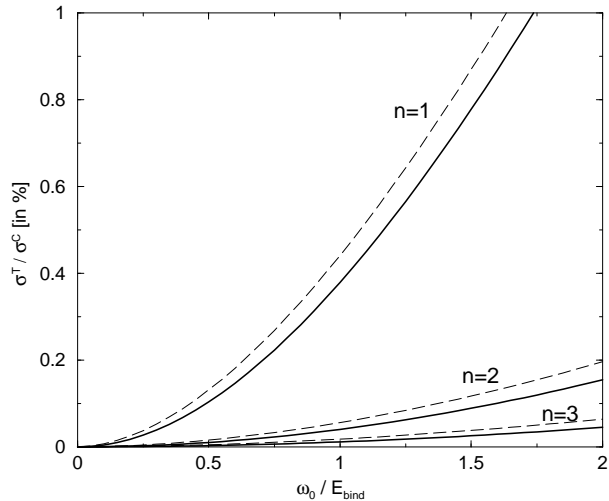}
\caption{Ratio (in per cent) between transverse electric and Coulomb 
contributions to the total cross section of pionium scattering off Ni 
at 5~GeV, as a function of $\omega_0$. Solid line: target-elastic 
process; dashed line: target-inclusive process. Pionium initial states 
(with $l_i=0$) are indicated in the figure.}
\label{trans}
\end{figure}
Figures~\ref{fig_contrib} and \ref{intbyshell} show clearly that the 
principal contributions to the cross sections come from the region 
$k_{\mathrm{TF}}<q_\perp < k_\Pi$. Thus the long-wavelength limit applies
for the pionium (but not for the atom).
Furthermore we noted that the relevant
excitation energy $\Delta_0$ is small and may safely be set to zero, and  
thus $W_{1,\mathrm{A}}=0$. Under
these circumstances the cross section due to the transverse part of the
current operator is given by \eref{sigmaelT} and \eref{sigmainelT}.
In \fref{trans} we show the ratio between the cross sections for the
transverse electric part ($T^\rme$) and the one for the Coulomb part 
($M^{\mathrm{C}}$), as a function of the average pionium excitation energy
$\omega_0$. Note that the ratio is given in per cent. The solid lines 
correspond to total cross sections for coherent scattering of pionium in 
various initial $s$-states as indicated in the figure. The dashed lines
correspond to the target-inclusive process (with $\Delta_0=0$). At the 
typical average pionium excitation energy $\omega_0=E_{\mathrm{bind}}$ the
transverse cross section $\sigma^{\mathrm{T}}$ for coherent scattering
amounts to 0.4\% of the
Coulomb part in the ground state, decreasing rapidly for initially excited
states of the pionium. 

\begin{figure}
\centering\includegraphics[height=5.5cm]{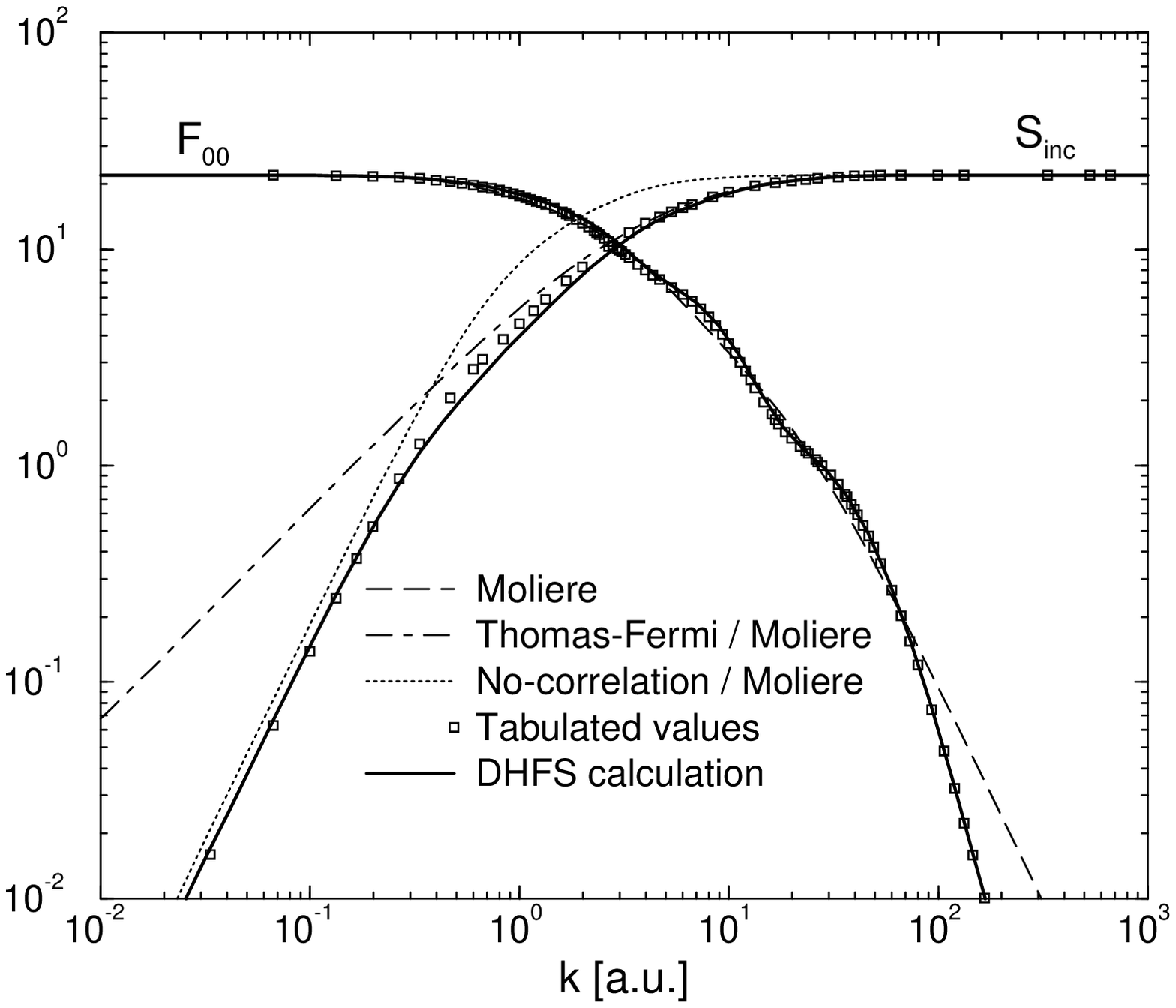}%
\quad\includegraphics[height=5.5cm]{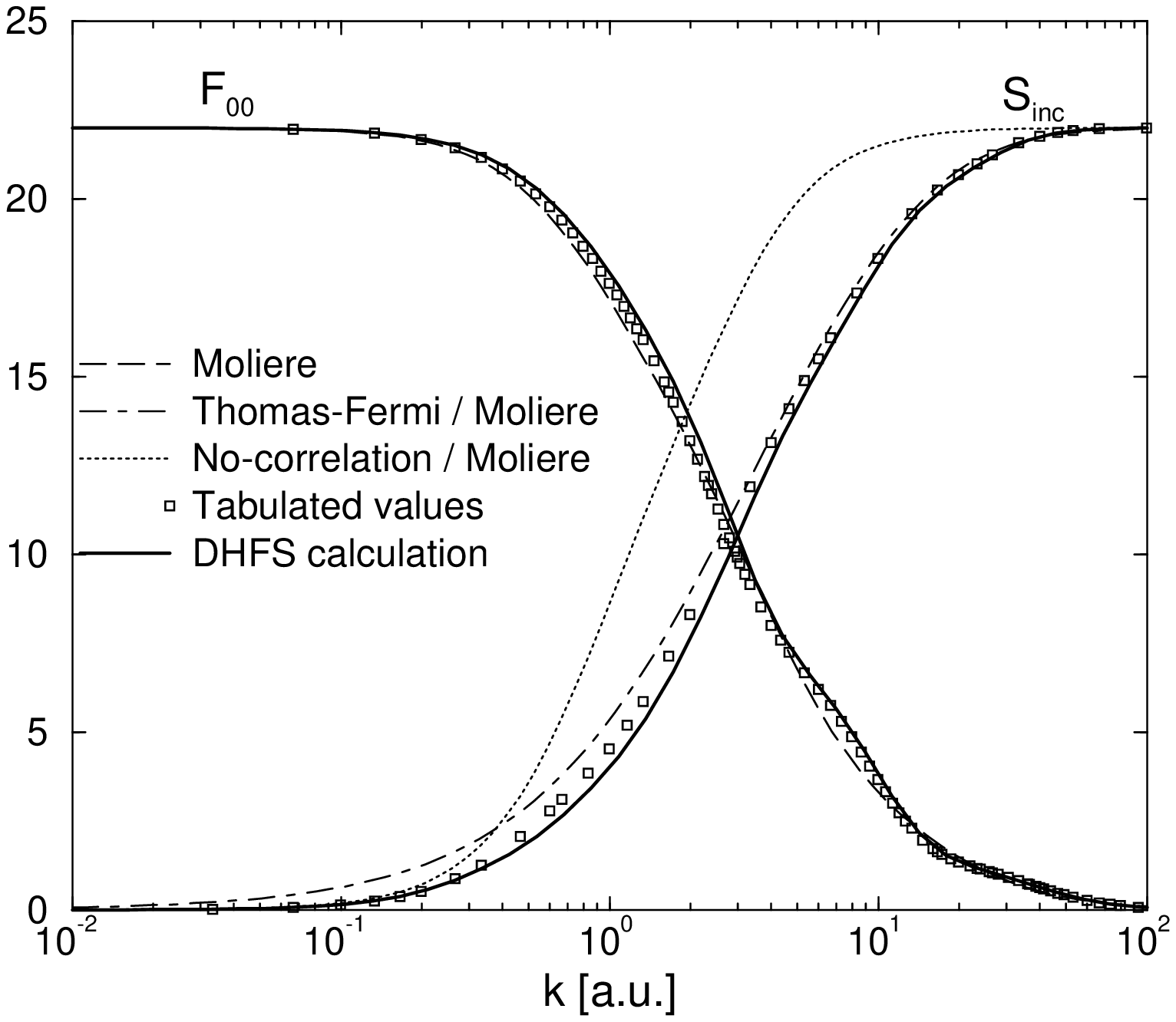}
\caption{Electronic part $F_{00}$ of the coherent atomic form factor and 
incoherent scattering function $S_{\mathrm{inc}}$ for Ti ($Z=22$). The 
asymptotic behavior is more easily
seen from the log-log diagram on the left. The range of relevance for the
cross section calculations is $0.1\le k\le 100$~a.u. For an explanation 
of the different models, see text.}
\label{FSinc}
\end{figure}

Finally \fref{FSinc} compares various models for $F_{00,\mathrm{A}}$
and $S_{\mathrm{inc,A}}$ over the range of $k$ values relevant for our
calculations of cross sections in the context of experiment DIRAC. The 
solid lines correspond to our calculations using \eref{F00DHFS} for the 
coherent form factor and \eref{SincDHFS} for the incoherent scattering 
function, respectively. We use relativistic Dirac orbitals in both cases.
The squares represent tabulated values from two compilations of
state-of-the-art Hartree-Fock calculations 
with various corrections (configuration interaction,
relativistic effects, a.s.o.). Note that while the tabulated results for
$F_{00}$ correspond to relativistic calculations \cite{HubbellF}, the 
tables for $S_{\mathrm{inc}}$ in \cite{HubbellS} contain only
non-relativistic results. The log-log representation in the left frame
serves merely to display the asymptotic behavior of the simple models
discussed in \sref{models}. Analyzing the dashed line corresponding
to \eref{Fmol} we note that the incorrect asymptotic fall-off for
$F_{00}$ at large $k$ poses no difficulties. 
In the right panel we see that between $5\le k\le 10$~a.u. 
the Thomas-Fermi-Moli\`ere model misses features of the atomic shell 
structure, but the resulting deviation from \eref{F00DHFS} is 
insignificant. Much more problematic are the crude approximations for
$S_{\mathrm{inc}}$. While the no-correlation limit \eref{nocorr}, using 
\eref{Fmol} and shown with the dotted line, 
increases with $k^2$ at small $k$ (as it should), it dramatically 
overestimates the scattering function in the most relevant range
between 0.1 and 100~a.u. On the other hand, the Thomas-Fermi model for 
incoherent scattering as developed by Heisenberg \cite{Heis}, with 
modifications \cite[App. B]{Tsai} for the use of Moli\`ere's 
approximation, is quite successful at $k\ge 5$~a.u., but it fails 
completely at smaller $k$.
\section{Conclusions}
\label{conclusion}
We have reviewed the formalism for incoherent atomic scattering. The 
basic expressions for the atomic form factors and scattering functions 
have then been evaluated in the framework of Dirac-Hartree-Fock-Slater
theory, i.e., using numerically determined electron orbitals. For 
comparison, both the form factor for coherent scattering, as well as the 
incoherent scattering function have been derived in simple analytical 
models based on the Thomas-Fermi model of the atom. Applying 
these different descriptions we performed detailed numerical studies
in the context of pionium scattering incoherently off the electrons of
various target atoms. Due to the much larger reduced mass of the pionium
system ($\mu_\Pi=m_\pi/2\approx 136.566 m_e$), the length and momentum
scales of the pionium and the target atom are very different. 

An investigation of the relevant momentum transfer $q_\perp$ revealed 
that the cross sections are dominated by the contributions from the 
region between the Thomas-Fermi momentum 
$k_{\mathrm{TF}}=Z^{1/3}/a_{\mathrm{Bohr}}$ for the atom, and the 
momentum scale of the pionium at $k_\Pi=\mu_\Pi/a_{\mathrm{Bohr}}$.
Under these circumstances the simple models for incoherent scattering
discussed in \sref{models} prove not sufficiently accurate for our
application in the context of pionium--atom scattering with a required
accuracy of 1\% or better. 
Whereas the analytical models discussed in \sref{models}
provide sufficiently accurate
\emph{coherent} form factors, the \emph{incoherent} contribution really
requires the more accurate treatment developed in \sref{calcDHFS}, 
despite its lesser importance (as compared to coherent scattering).
Only an explicit DHFS calculation can provide satisfactory scattering 
functions. 
Since the target materials employed in experiment DIRAC will be
as heavy as Pt, our calculation with relativistic orbitals is clearly more 
appropriate than the non-relativistic results tabulated in \cite{HubbellS}.

From our detailed discussion of the $q_\perp$-dependence of the 
integrand for the cross section we also conclude that the loosely bound 
outer shell electrons dominate the target-inelastic cross sections. 
Their contribution to the integrand $\rmd\sigma /\rmd q_\perp$ covers a 
much larger range of $q_\perp$ values than the one corresponding to inner 
shell electrons. Following this argument further, 
free electrons would show a behavior rather similar to
the one of the outer shells, their contribution to the cross section would
stretch even further down to smaller $q_\perp$. However, the cross section
hardly depends on this modification at very small $q_\perp$.
Our calculation therefore applies equally well to quasi-free electrons and 
to electrons in the conduction band. Thus solid state effects and chemistry
need not be considered explicitly as they prove 
irrelevant for calculations pertaining to experiment DIRAC. The same 
conclusion had been inferred in our earlier study 
\cite{HHHTV99} based on an analysis of the relevant impact
parameter range. Recalling 
that target-inelastic scattering constitutes merely a correction of
order $1/Z$ of the atomic part, variations on the order of a few per cent
in the incoherent scattering cross sections are insignificant. 

These findings are confirmed in our analysis of the dependence of the 
cross sections on the average excitation energies for pionium and atom. 
Our calculation of incoherent scattering contributions resolved according 
to individual target electron shells demonstrates that the average 
excitation energy for the atom may safely be set to zero, $\Delta_0=0$. 
This information which is crucial for the closure approximation cannot be 
extracted from the scattering functions in \cite{HubbellS}.
For the pionium, on the other hand, a non-vanishing excitation energy in 
the amount of the binding energy is needed when calculating total cross 
sections in the closure approximation 
in order to get agreement between
the closure result and the result obtained by explicitly summing the partial
cross sections over all final states.

Earlier investigations on pionium--atom interaction 
\cite{Afanasyev,HHHTV99}
invoked properties of the hydrogen-like pionium system to obtain crude
estimates for the magnitude of the magnetic terms so far neglected in 
this interaction. From the non-relativistic wavefunctions for the 
pionium, its internal
velocity is of the order of $v/c\approx\alpha/2$. Thus magnetic terms 
are believed to be small of this order. 
We established a much better justified estimate for the magnetic terms 
(actually for the transverse part of the current) in the long-wavelength
limit. Our investigation showed that this limit applies very well for 
the pionium, whereas it does not apply for the atom. However, the
transverse current does not contribute in the elastic case on the atom 
side, and for the inelastic case it is suppressed even more than the 
corresponding term on the pionium side because the relevant atomic 
excitation energies (of the outer shells) are smaller than those of the 
pionium.
\ack
We would like to thank R.D. Viollier, J. Schacher, L. Nemenov, and L.
Afanasyev for stimulating discussions on the subject of this article.
Special thanks are due to Zlatko Halabuka whose efficient and accurate
integration routine proved instrumental in the calculation of atomic 
form factors and scattering functions. 
We also wish to thank an unknown
referee for fruitful and stimulating comments.

\end{document}